\documentclass[draft,a4paper]{article}

\usepackage[applemac]{inputenc}
\usepackage{latexsym}
\usepackage{amssymb}
\usepackage{amsmath}
\usepackage{ifthen}
\usepackage{enumerate}
\newcommand{\set}[2]{\left\{#1\mathrel{\left|\vphantom{#1}\vphantom{#2}\right.}#2\right\}}
\newcommand{\oneset}[1]{\left\{\mathinner{#1}\right\}}
\newcommand{\smallset}[1]{\left\{\mathinner{#1}\right\}}

\newcommand{\abs}[1]{\left|\mathinner{#1}\right|}

\newcommand{\gR}{\mathrel{\mathcal{R}}}

\newcommand{\gL}{\mathrel{\mathcal{L}}}

\newcommand{\gH}{\mathrel{\mathcal{H}}}

\newcommand{\gJ}{\mathrel{\mathcal{J}}}

\newcommand{\gRop}{\mathbin{\gR}}
\newcommand{\gLop}{\mathbin{\gL}}
\newcommand{\gJop}{\mathbin{\gJ}}
\newcommand{\gHop}{\mathbin{\gH}}

\newcommand{\SU}{\mathbin{\mathsf{S\kern-0.1emU}}}

\renewcommand{\phi}{\varphi}

\newenvironment{proof}[1][]{\pagebreak[3]\noindent\textit{Proof\ifthenelse{\equal{#1}{}}{}{ (#1)}: }}{\pagebreak[3]\medskip}
\newtheorem{expl}{Example}[section]

\newcommand{\qed}{\hspace*{\fill}$\Box$}

\begin{document}

\title{A Proof of the Factorization Forest Theorem}

\author{
  Manfred Kuf\-leitner$^{1,2}$ \\
  {\footnotesize {${}^1$}LaBRI, Universit{\'e} de Bordeaux and CNRS} \\[-1.7mm]
  {\footnotesize 351, Cours de la Lib{\'e}ration} \\[-1.7mm]
  {\footnotesize F-33405 Talence cedex, France} \\
  {\footnotesize  {${}^2$}FMI, Universit{\"a}t Stuttgart} \\[-1.7mm]
  {\footnotesize  Universit{\"a}ts\-str.~38} \\[-1.7mm]
  {\footnotesize  D-70569 Stuttgart, Germany}
}

\date{}

\maketitle

\begin{abstract}
  We show that for every homomorphism $\Gamma^+ \to S$ where $S$ is a
  finite semigroup there exists a factorization forest of height $\leq
  3 \abs{S}$. The proof is based on Green's relations.
\end{abstract}

\section{Introduction}\label{sec:intro}

Factorization forests where introduced by Simon \cite{sim87,sim90}. An
important property of finite semigroups is that they admit
factorization forests of finite height. This fact is called the
Factorization Forest Theorem.  It can be considered as an Ramsey-type
property of finite semigroups.  There exists different proofs of this
fact of different difficulty and with different bounds on the
height. The first proof of the Factorization Forest Theorem is due to
Simon \cite{sim90}. He showed that for every finite semigroup $S$
there exists a factorization forest of height $\leq 9 \abs{S}$. The
proof relies on several different techniques. It uses graph colorings,
Green's relations, and a decomposition technique inspired by the
Rees-Suschkewitsch Theorem on completely $0$-simple semigroups.  In
\cite{sim92} Simon gave a simplified proof relying on the Krohn-Rhodes
decomposition. The bound shown is $2^{\abs{S}+1}-2$. A concise proof
has been given by Chalopin and Leung \cite{ChaLeu04}.  The proof
relies on Green's relations and yields the bound $7 \abs{S}$ on the
height.  Independently of this work, Colcombet has also shown a bound
of $3 \abs{S}$ for the height of factorization forests
\cite{col07fct}.  He uses a generalization of the Factorization Forest
Theorem in terms of \emph{Ramseyan splits}. The proof also relies on
Green's relations. A variant of our proof for the special case of
aperiodic monoids has been shown in \cite{dk07dlt} with a bound of $3
\abs{S}$. The main benefit of that proof is that is uses very little
machinery. The proof in this paper can be seen as an extension of that
proof. The main tool are again Green's relations. We only require
basic results from the theory of finite semigroups which can be found
in standard textbooks such as \cite{pin86}.

A lower bound of $\abs{S}$ was shown for rectangular bands in
\cite{sim88} and also in \cite{ChaLeu04}. The same bound has also been
shown for groups \cite{ChaLeu04}. Therefore, the upper bound of $3
\abs{S}$ reduces the gap between the lower and the upper bound.

\section{The Factorization Forest Theorem}

Let $S$ be a finite semigroup.  A \emph{factorization forest} of a
homomorphism $\varphi : \Gamma^+ \to S$ is a function $d$ which maps
every word $w$ with length $\abs{w} \geq 2$ to a factorization $d(w) =
(w_1, \ldots, w_{n})$ of $w = w_1 \cdots w_{n}$ with $n \geq 2$ and
$w_{i} \in \Gamma^+$ and such that $n \geq 3$ implies $\varphi(w_1) =
\cdots = \varphi(w_{n})$ is idempotent in $S$. The \emph{height} $h$
of a word $w$ is defined as
\begin{equation*}
  h(w) = 
  \begin{cases}
    0 & \text{if } \abs{w} \leq 1 \\
    1 + \max\oneset{h(w_1), \ldots, h(w_{n})} & 
    \text{if } d(w) = (w_1, \ldots, w_{n})
  \end{cases}
\end{equation*}
We call the tree defined by the ``branching'' $d$ for the word $w$ the
\emph{factorization tree} of $w$.  The height $h(w)$ is the height of
this tree. The height of a factorization forest is the supremum over
the heights of all words.

\bigskip

\noindent
\textbf{Factoriztion Forest Theorem (Simon \cite{sim90}).}  
\emph{Let $S$ be a finite monoid. Every homomorphism $\varphi : \Gamma^+ \to
S$ has a factorization forest of height $\leq 3 \abs{S}$.}

\bigskip

\begin{proof}
  Let $[w] = \varphi(w)$.  We show that for every $w \in \Gamma^+$
  there exists a factorization tree of height $h(w) \leq 3 \abs{\set{x
      \in S}{[w] \leq_{\gJ} x}}$. First, we perform an induction on
  the cardinality of the set $\set{x \in S}{[w] \leq_{\gJ} x}$; then
  within one $\gJ$-class we refine this parameter. Let $w \in
  \Gamma^+$ with $\abs{w} \geq 2$. Then $w$ has a unique factorization
  \begin{equation*}
    w = w_0 a_1 w_1 \cdots a_m w_m
  \end{equation*}
  with $a_i \in \Gamma$ and $w_i \in \Gamma^*$ satisfying the
  following two conditions:
  \begin{equation*}
    \forall\, 1 \leq i \leq m \colon [a_i w_i] \gJop [w]
    \;\ \text{ and } \;\ 
    \forall\, 0 \leq i \leq m \colon w_i = \varepsilon 
    \; \vee \; [w] <_{\gJ} [w_i]
  \end{equation*}
  The idea is that we successively choose $a_i w_i \in \Gamma^+$ from
  right to left to be the shortest non-empty word such that $[a_i w_i]
  \gJop [w]$.  Let $w_i' = a_i w_i$ for $1 \leq i \leq m$.  For each
  $1 \leq i < m$ define a pair $(L_i,R_i)$ where $L_i$ is the
  $\gL$-class of $[w_i']$ and $R_i$ is the $\gR$-class of
  $[w_{i+1}']$. Every such pair represents an $\gH$-class within the
  $\gJ$-class of $[w]$. All $\gH$-classes within this $\gJ$-class
  contain the same number $n$ of elements. Let
  \begin{equation*}
    h'(w) \;=\; h(w) - 3 \cdot \abs{\set{x \in S}{[w] <_{\gJ} x}}
  \end{equation*}
  We can think of $h'$ as the height of a tree where we additionally
  allow words $v \in \Gamma^+$ with $[w] <_{\gJ} [v]$ as leafs.
  Within the $\gJ$-class of $w$ we perform an induction on the
  cardinality of the set $\set{(L_i,R_i)}{1 \leq i < m}$ in order to
  show
  \begin{equation*}
    h'(w) \;\leq\;
    3 n \cdot \abs{\set{(L_i,R_i)}{1 \leq i < m}}
  \end{equation*}
  Since $n \cdot \abs{\set{(L_i,R_i)}{1 \leq i < m}} \,\leq\,
  \abs{\set{x \in S}{[w] \,\gJop\, x}}$ this yields the desired bound
  for the height $h(w)$. If every pair $(L,R)$ occurs at most twice
  then we have $m-1 \leq 2 \cdot \abs{\set{(L_i,R_i)}{1 \leq i < m}}$.
  We define a factorization tree for $w$ by
  \begin{alignat*}{2}
    d(w) &= (w_0 w_1',\,w_2' \cdots w_m') && \\
    d(w_0 w_1') &= (w_0, w_1') && \\
    d(w_i' \cdots w_m') &= (w_i', w_{i+1}' \cdots w_m')
    &&\qquad \text{for } 2 \leq i < m \\
    d(w_i') &= (a_i, w_i)
    &&\qquad \text{for } 1 \leq i \leq m
  \end{alignat*}
  Since $[w] <_{\gJ} [w_i]$, by induction every $w_i$ has a
  factorization tree of height $h(w_i) \;<\; 3 \abs{\set{x}{[w_i]
      \leq_{\gJ} x}} \;\leq\; 3 \abs{\set{x}{[w] <_{\gJ} x}}$.  This
  yields:
  \begin{align*}
    h'(w) \;\leq\; m \;\leq\;
    3 n \cdot \abs{\set{(L_i,R_i)}{1 \leq i < m}}
  \end{align*}
  Note that the height does not increase if some of the $w_i$ are
  empty.  Now suppose there exists a pair $(L,R) \in \set{(L_i,R_i)}{1
    \leq i < m}$ occurring (at least) three times. Let $i_0 < \cdots <
  i_k$ be the sequence of all positions with $(L,R)=(L_{i_j},
  R_{i_j})$. By construction we have $k \geq 2$.  Let $\widehat{w_j} =
  w_{i_{j-1} + 1}' \cdots w_{i_j}'$ for $1 \leq j \leq k$. For all $1
  \leq j \leq \ell \leq k$ we have
  \begin{itemize}
  \item $[\widehat{w_j} \cdots \widehat{w_{\ell}}] 
    \,\leq_{\gL}\, [w_{i_{\ell}}'] \,\gLop\, [w_{i_{0}}']$.
  \item $[\widehat{w_j} \cdots \widehat{w_{\ell}}]
    \,\leq_{\gR}\, [w_{i_{j-1}+1}'] \,\gRop\, [w_{i_{0}+1}']$.
  \item $[w_{i_{\ell}}'] 
    \,\leq_{\gJ}\, [\widehat{w_j} \cdots \widehat{w_{\ell}}] 
    \,\leq_{\gJ}\, [w]
    \,\gJop\, [w_{i_{\ell}}'] \,\gJop\, [w_{i_{0}}']
    \,\gJop\, [w_{i_{0}+1}']$ by assumption on the
    factorization.
  \end{itemize}
  Thus for all $1 \leq j \leq \ell \leq k$ and $1 \leq j' \leq \ell'
  \leq k$ we get
  \begin{itemize}
  \item $[\widehat{w_j} \cdots \widehat{w_{\ell}}] \,\gLop\,
    \,[w_{i_{1}}'] \,\gLop\, [\widehat{w_{j'}} \cdots
    \widehat{w_{\ell'}}]$ \ and
  \item $[\widehat{w_j} \cdots \widehat{w_{\ell}}] \,\gRop\,
    [w_{i_{1}+1}'] \,\gRop\, [\widehat{w_{j'}} \cdots
    \widehat{w_{\ell'}}]$ \ and therefore
  \item $[\widehat{w_j} \cdots \widehat{w_{\ell}}] \,\gHop\, 
    [\widehat{w_{j'}} \cdots \widehat{w_{\ell'}}]$
  \end{itemize}
  Therefore, all $[\widehat{w_j}]$ denote elements in the same
  $\gH$-class $H$ and since $k \geq 2$ the class $H$ is a group.  We
  consider the following set of elements in $H$ induced by proper
  prefixes
  \begin{equation*}
    P(\widehat{w_1}\cdots\widehat{w_k})
    = \set{[\widehat{w_1} \cdots \widehat{w_j}]}{1 \leq j < k}
  \end{equation*}
  For the pair $(L,R)$ we show by induction on
  $\abs{P(\widehat{w_1}\cdots\widehat{w_k})}$ that
  \begin{align*}
    h'(w) \ \leq \ 3 \abs{P(\widehat{w_1}\cdots\widehat{w_k})} \ +\ 3 n
    \abs{\set{(L_i,R_i)}{1 \leq i < m}}
  \end{align*}
  Suppose every element $x \in P(\widehat{w_1} \cdots \widehat{w_k})
  \subseteq H$ occurs at most twice.  Then $k-1 \leq 2
  \abs{P(\widehat{w_1}\cdots\widehat{w_k})}$. We construct the
  following factorization tree for $w$:
  \begin{align*}
    d(w) &= (w_0 w_1' \cdots w_{i_1}',\; w_{i_1 + 1}' \cdots w_m') \\
    d(w_0 w_1' \cdots w_{i_1}') &= (w_0 w_1' \cdots w_{i_0}',\; 
    \widehat{w_1}) \\
    d(w_{i_0 + 1}' \cdots w_m') 
    &= (\widehat{w_2} \cdots \widehat{w_k},\; w_{i_k + 1}' \cdots w_m') \\
    d(\widehat{w_i} \cdots \widehat{w_k}) &
    = (\widehat{w_i},\; \widehat{w_{i+1}} \cdots \widehat{w_k})
    \qquad\quad \text{for } 2 \leq i < k
  \end{align*}
  By induction on the number of pairs $(L_i,R_i)$ there exist
  factorization trees for the words $w_0 w_1' \cdots w_{i_0}'$,
  $w_{i_k + 1}' \cdots w_m'$, and all $\widehat{w_{i}}$ of height
  \begin{equation*}
    \leq \ 3 n \abs{\set{(L_i,R_i)}{1 \leq i < m} 
      \setminus \smallset{(L,R)}}
    \;+\; 3 \abs{\set{x}{[w] <_{\gJ} x}} 
  \end{equation*}
  This yields
  \begin{align*}
    h'(w) - 3 n \abs{\set{(L_i,R_i)}{1 \leq i < m}} \ \leq \ k 
    \ \leq \ 
    3 \abs{P(\widehat{w_1}\cdots\widehat{w_k})}
  \end{align*}
  Now suppose there exists an element $x \in P(\widehat{w_1} \cdots
  \widehat{w_k}) \subseteq H$ that occurs at least three times. Let
  $j_0 < \cdots < j_t$ be the sequence of all positions with $x =
  [\widehat{w_1} \cdots \widehat{w_{j_i}}]$. By construction we have
  $t \geq 2$.  It follows that $[\widehat{w_{j_i + 1}} \cdots
  \widehat{w_{j_{i+1}}}] = e=e^2$ where $e$ is the neutral element of
  the group $H$. Let $v_i = \widehat{w_{j_{i-1} + 1}} \cdots
  \widehat{w_{j_{i}}}$ for $1 \leq i \leq t$.  We construct the
  following factorization tree for $w$:
  \begin{align*}
    d(w) &=
    (w_0 \cdots \widehat{w_{i_0}},\; \widehat{w_{i_0 + 1}} \cdots w_m') \\
    d(\widehat{w_{i_0 + 1}} \cdots w_m') &= 
    (v_1 \cdots v_t,\; \widehat{w_{i_t + 1}} \cdots w_m') \\
    d(v_1 \cdots v_t) &= (v_1, \ldots, v_t)
  \end{align*}
  We have $x \in P(\widehat{w_1}\cdots\widehat{w_k}) \setminus
  P(\widehat{w_1} \cdots \widehat{w_{i_0}})$ and $x
  P(\widehat{w_{j_{i-1} + 1}} \cdots \widehat{w_{j_{i}}}) \subseteq
  P(\widehat{w_1}\cdots\widehat{w_k})$ but $x \not\in x
  P(\widehat{w_{j_{i-1} + 1}} \cdots \widehat{w_{j_{i}}})$. Hence, by
  induction on the cardinality of the prefix sets, there exist
  factorization forests for $w_0 \cdots \widehat{w_{i_0}}$,
  $\widehat{w_{i_t + 1}} \cdots w_m'$ and the $v_i$ of height
  \begin{align*}
    \leq \ 
    & 3 \abs{P(\widehat{w_1} \cdots \widehat{w_k})} \; - \; 3 \\
    & {+}\ 3 n \abs{\set{(L_i,R_i)}{1 \leq i < m} 
      \setminus \smallset{(L,R)}} \\
    & {+}\ 3 \abs{\set{x}{[w] <_{\gJ} x}}
  \end{align*}
  This yields a factorization tree for $w$ with the desired height
  bound. \qed
\end{proof}

\medskip

\noindent
\textbf{Acknowledgement.}  I would like to thank Volker Diekert for
numerous discussions on this topic.


\begin{thebibliography}{1}

\bibitem{ChaLeu04}
J{\'e}r{\'e}mie Chalopin and Hing Leung.
\newblock On factorization forests of finite height.
\newblock {\em Theoretical Computer Science}, 310(1-3):489--499, 2004.

\bibitem{col07fct}
Thomas Colcombet.
\newblock Factorisation forests for infinite words.
\newblock In Erzs{\'e}bet Csuhaj-Varj{\'u} and Zolt{\'a}n {\'E}sik, editors,
  {\em Fundamentals of Computation Theory, 16th International Symposium, {FCT}
  2007, Budapest, Hungary, August 27-30, 2007, Proceedings}, volume 4639 of
  {\em Lecture Notes in Computer Science}, pages 226--237. Springer-Verlag,
  2007.

\bibitem{dk07dlt}
Volker Diekert and Manfred Kuf\-leitner.
\newblock On first-order fragments for words and {M}azurkiewicz traces: {A}
  survey.
\newblock In Tero Harju, Juhani Karhum{\"{a}}ki, and Arto Lepist{\"{o}},
  editors, {\em Developments in Language Theory, 11th International Conference,
  DLT 2007, Turku, Finland, July 3-6, 2007, Proceedings}, volume 4588 of {\em
  Lecture Notes in Computer Science}, pages 1--19. Springer-Verlag, 2007.

\bibitem{pin86}
Jean-{\'{E}}ric Pin.
\newblock {\em {Varieties of Formal Languages}}.
\newblock North Oxford Academic, London, 1986.

\bibitem{sim87}
Imre Simon.
\newblock Factorization forests of finite height.
\newblock Technical Report 87-73, Laboratoire d'Informatique Th{\'e}orique et
  Programmation, Paris, 1987.

\bibitem{sim88}
Imre Simon.
\newblock Properties of factorization forests.
\newblock In Jean-{\'E}ric Pin, editor, {\em Formal Properties of Finite
  Automata and Applications: {LITP} Spring School on Theoretical Computer
  Science}, volume 386 of {\em Lecture Notes in Computer Science}, pages
  65--72. Springer-Verlag, 1988.

\bibitem{sim90}
Imre Simon.
\newblock Factorization forests of finite height.
\newblock {\em Theoretical Computer Science}, 72(1):65--94, 1990.

\bibitem{sim92}
Imre Simon.
\newblock A short proof of the factorization forest theorem.
\newblock In Maurice Nivat and Andreas Podelski, editors, {\em Tree Automata
  and Languages}, pages 433--438. Elsevier, 1992.

\end{thebibliography}
\end{document}